\documentclass[showpacs,preprintnumbers,amsmath,amssymb]{revtex4}

\usepackage{graphicx}
\newcommand{\Gch}{\check{G}}
\newcommand{\Ght}{\hat{G}}
\newcommand{\Sch}{\check{\Sigma}}
\newcommand{\Shts}{\hat{\Sigma}_{S\pm}}
\newcommand{\Shtf}{\hat{\Sigma}_{F\pm}}
\newcommand{\gch}{\check{g}}

\newcommand{\ghts}{\hat{g}_{S\pm}}
\newcommand{\ghtf}{\hat{g}_{F\pm}}
\newcommand{\fhts}{\hat{f}_{S\pm}}
\newcommand{\fhtf}{\hat{f}_{F\pm}}
\begin{document}

\title{Inverse proximity effect in superconductor-ferromagnet structures: From the ballistic to the diffusive limit}
\author{F. S. Bergeret}
\author{A. Levy Yeyati}
\author{A. Mart\'{i}n-Rodero}
\affiliation{Departamento de F\'{i}sica Te\'{o}rica de la Materia
Condensada C-V, Universidad Aut\'{o}noma de Madrid, E-28049
Madrid, Spain}

\begin{abstract}
The inverse proximity effect, i.e. the induction of a magnetic
moment in the superconductor in superconductor-ferromagnet (S/F)
junctions is studied theoretically. We present a microscopic
approach which combines a model Hamiltonian with elements of the
well established quasiclassical theory. With its help we  study
systems with arbitrary degree of disorder, interface transparency
and thickness of the layers. In the diffusive limit we recover the
result of  previous works: the direction of the  induced
magnetization $M$ is opposite to the one of the F layer. However,
we show that in the ballistic case the sign of $M$  may be
positive or negative depending on the quality of the interface and
thickness of the layers. { We show that, regardless of its sign,
the penetration length of the magnetic moment into the
superconductor is of the order of the superconductor coherence
length, which demonstrates that the effect has a superconducting
origin.}
\end{abstract}

\maketitle

\section{Introduction}

The proximity effect in multilayered structures consisting of
superconductors and normal metals has been studied intensively in
the last decades. Since the sixties it is well known that
superconducting correlations can penetrate into a normal region
changing the physical properties of the latter \cite{de_gennes}.
More recently special interest has been paid to the study of the
proximity effect in superconductor-ferromagnet structures. The
presence of an exchange field acting on the spin of the electrons
leads to interesting new physics, as for example oscillations in
the superconducting critical temperature and in the density of
states (DoS) of the ferromagnet as a functions of its thickness
\cite{radovic1,lazar,kontos01}. The interplay between
superconductivity and magnetism  also leads to  a change of the
sign of the Josephson current in S/F/S structures
\cite{buzdin,ryazanov} (for a review see \cite{buzdin_rev}). On
the other hand, recent theoretical studies have focused on the
{\it inverse} proximity effect, i.e. the induction of a magnetic
moment in a superconductor in contact with a ferromagnet
\cite{BVE_inverse,halterman_04,buzdin05,kriv}. This phenomenon has
been analyzed in Refs. \cite{BVE_inverse,kriv} for diffusive  S/F
junctions within the quasiclassical Green functions formalism. The
results of Ref. \cite{BVE_inverse} demonstrate that the direction
of the induced magnetic moment is opposite to the one in the
ferromagnet. The authors of Ref.\cite{BVE_inverse} also shown that
this effect is related to the induction of a triplet component
(with the zero projection $S_z=0$ of the total spin) in the
pairing amplitude.

The interest in the inverse proximity effect has been increased
with the development of experimental techniques like neutron
reflectometry \cite{stahn} and muon spin rotation \cite{muon}
which allow to determine accurately the spatial distribution of
magnetic moments. For example experiments on multilayered system
consisting of the high $T_C$ superconductor YBa$_2$Cu$_3$O$_7$ and
the ferromagnet L$_{2/3}$C$_{1/3}$MnO$_3$ layers have shown an
induced magnetic moment in the S layers with a sign opposite to
the magnetization of the ferromagnet \cite{stahn}. However, a
conclusive explanation of this experiment has not been given yet.

More recently the inverse proximity effect was studied
theoretically in atomic-scale F/S/F trilayers \cite{buzdin05}. It
was shown that if the magnetization of the F layers points in the
same direction, the S layer acquires a spin polarization parallel
to the one in the F layers. At first glance this result
contradicts the one of Ref.\cite{BVE_inverse}, but until now no
detailed comparison between these two works has been done.  For
completeness we mention that in Ref.\cite{halterman_04} the
induced magnetization in a clean S/F bilayer was computed for
highly polarized ferromagnets and a leakage of the magnetic moment
into the superconductor over atomic distances was obtained. {In
Ref.\cite{tokuyasu} the induced magnetic moment in a clean
superconductor in contact with a magnetic wall was studied. }

In this work we present a microscopic model which allows to study
the inverse proximity effect for systems with an arbitrary degree
of disorder, interface transparency and layer thickness. The model
is based on the Hamiltonian approach of Refs.
\cite{josephson,cuevas,vecino}. Within this model we show how
disorder can be introduced by an appropriate averaging procedure
in such a way as to recover the quasiclassical theory in the
diffusive limit.

The main results of the paper can be summarized as follows: For an
atomic-size bilayer in the pure ballistic limit, in which the
component of the momentum parallel to the plane of the layers is
conserved, the induced magnetic moment has a positive sign in
accordance with the results of Ref.\cite{buzdin05}. However this
is only valid for low values of the interface transmission
coefficient $\tau$.  When $\tau$ is of the order of unity the sign
of the induced magnetization may be negative. These results can be
understood in terms of the behavior of the bilayer Andreev states
as discussed in Sect. III. When disorder is included and the
interface becomes diffusive  we recover the result obtained in
Ref.\cite{BVE_inverse}, according to which the induced magnetic
moment in S is negative regardless of the interface transparency.
We also consider the intermediate situation and show that one can
go smoothly from the ballistic to the diffusive regime. In section
IV we consider the case of a S layer with thickness much larger
than the superconducting coherence length $\xi_S$. We show that
the sign of the induced magnetic  moment in the superconductor
depends on the thickness of the F layers as well as on the
transmission of the interface. We also show that the penetration
of the magnetic moment into the superconductor is in all cases of
the order of $\xi_S$. This demonstrates that the effect is
governed by superconducting correlations and is not just a
"magnetization leakage" over atomic distances into the S region.

{The rest of the paper is organized as follows: in Sect. II we
introduce the model Hamiltonian. Details of the model are
presented in the appendix, where we  also show the equivalence
between our model and  the quasiclassical approach in the
appropriate limit.} In section III we study the inverse proximity
effect in a S/F bilayer of atomic-size, while in section IV we
consider layers of arbitrary thickness. Finally section V is
devoted to the concluding remarks.

\section{Theoretical approach}

In order to describe a S/F layered structure we use a lattice
model Hamiltonian as in Ref. \cite{vecino}. This can be written as
$H=H_S+H_F+H_{SF}$ where
\begin{eqnarray}
H_S&=&\sum_{i,\sigma=\pm}\epsilon_{Si}a^+_{i,\sigma}a_{i,\sigma}+\sum_{i}\Delta_i(a^+_{i,+}a^+_{i,-}
+a_{i,-}a_{i,+})+
\sum_{<i,j>\sigma=\pm}t_{Sij}(a^+_{i,\sigma}a_{j,\sigma}+a^+_{i,\sigma}a_{j,\sigma}) \label{hs}\\
H_F&=&\sum_{i,\sigma=\pm}\epsilon_{Fi}a^+_{i,\sigma}a_{i,\sigma}-\sum_{i}h_i(a^+_{i,+}a_{i,+}-a^+_{i,-}a_{i,-})+
\sum_{<i,j>,\sigma=\pm}t_{Fij}(a^+_{i,\sigma}a_{j,\sigma}+a^+_{j,\sigma}a_{i,\sigma})\;,\label{hf}
\end{eqnarray}
are the terms describing the uncoupled superconducting and ferromagnetic regions.
Indexes $i,j$ label different sites along the system. The coefficients
$t_{ij}$ are the hopping amplitudes between neighboring sites. The site
energies $\epsilon_i$ are measured from the Fermi
level. $\Delta_i$ is the local order parameter and $h_i$ is the
local exchange field in the ferromagnet. Throughout this paper we assume
that $\Delta$ ($h$) is a given parameter which is constant and finite in
the superconductor (ferromagnet) and is equal to zero in the ferromagnet (superconductor).

Finally $H_{SF}$ is the term which describes the coupling between
the superconducting and ferromagnetic regions, which is given by
\begin{equation}
H_{SF}=\sum_{i\in S,j\in
F,\sigma=\pm}t_{SF,ij}\left(a^+_{i,\sigma}a_{j,\sigma}+
a^+_{j,\sigma}a_{i,\sigma}\right)\label{hsf}\; .
\end{equation}
Written in this way the Hamiltonian  can be used to describe many
different physical situations. In particular we would like to
consider not only ballistic but also diffusive systems. One can
include disorder either in the diagonal terms, i.e. introducing a
random correction $\delta\epsilon_i$ for the site energies, or
$\delta t_{ij}$ for the hopping terms. One can also model a
disordered interface by adding a random term $\delta t_{ij}$ to
the hopping elements of Eq. (\ref{hsf}).

The Green functions $\check{G}_n(k)$, which are 4$\times$4
matrices in the particle-hole (Nambu) $\otimes$ spin space should
be determined from the Dyson equation (see for example Refs.
\cite{vecino,BVE_josephson})
\begin{equation}
\Gch_{S(F)}=\gch_{S(F)}+\gch_{S(F)}(k)\Sch_{S(F)}\Gch_{S(F)}\;
,\label{dyson}
\end{equation}
where $\gch_{S(F)}$ is the Green function  at the interface on the
S(F) side for the uncoupled regions.  As discussed in the appendix
\ref{appA}  the self-energy $\Sch_{S(F)}$ in the ballistic case is
given by
\begin{equation}\label{Sigma_b}
    \Sch_{S(F)}^b(k)=t_{SF}^2\tau_3\check{g}_{F(S)}(k)\tau_3\;,
\end{equation}
while in the case of a diffusive interface has the form
\begin{equation}\label{Sigma_d}
\Sch_{S(F)}^d=\delta
 t^2\tau_3\langle \hat{G}_{F(S)\pm}\rangle\tau_3\; ,
\end{equation}
where $<...>$ denotes integration over parallel momentum $k$. The
problem is then reduced to a closed set of self-consistent
equations which can be solved as discussed in the next sections.
\begin{figure}[h]
 \includegraphics[ angle=-90, scale=0.4]{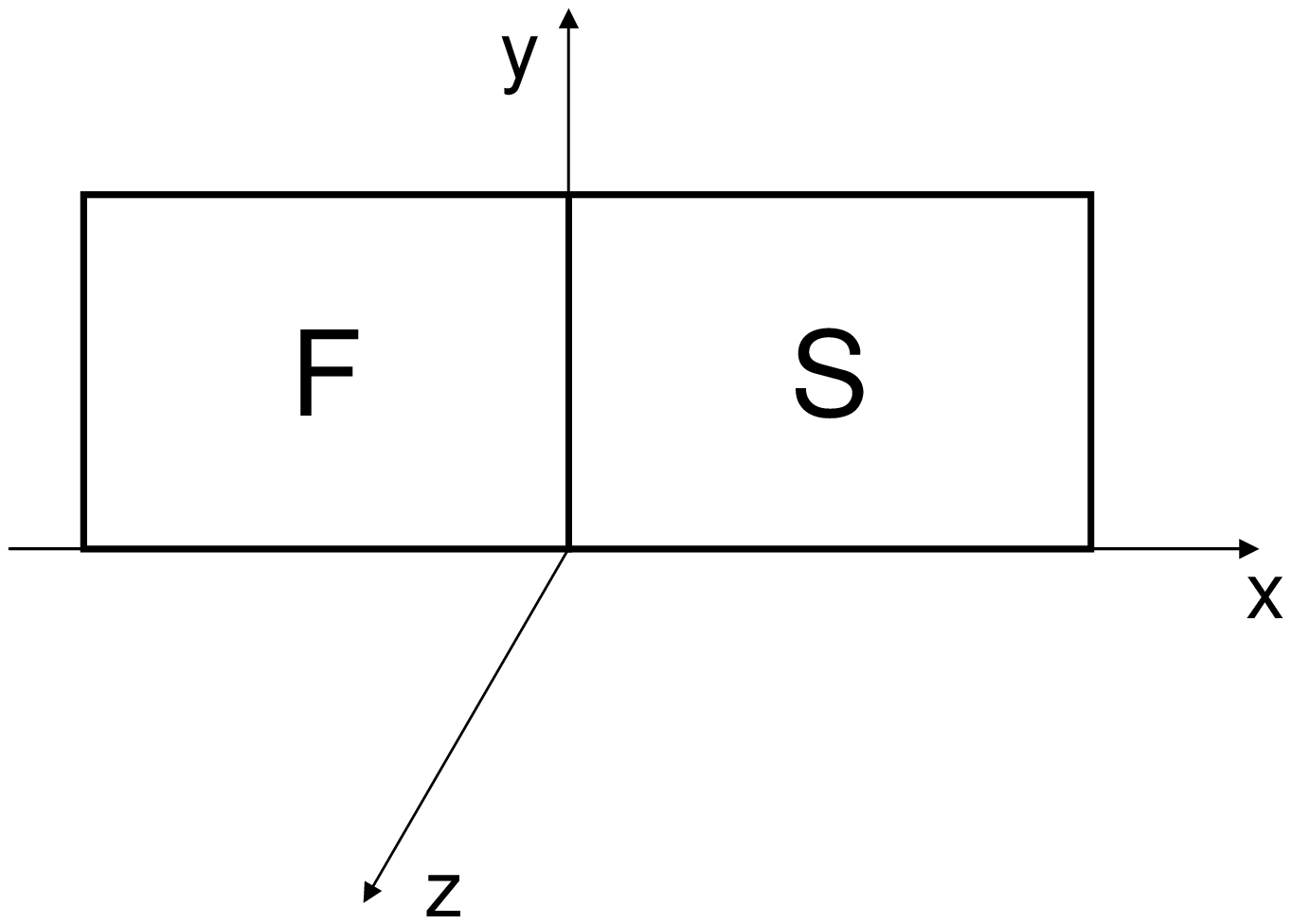}
 \caption{Geometry of the S/F structure.\label{geometry}}
\end{figure}

With the help of this model we can also describe a diffusive metal
by assuming that also the hopping terms between the planes inside
the S and F layers are randomly distributed. As we show in the
appendix \ref{appB}, after averaging over disorder our model leads
to a discretized version of  the well known Usadel equation
\cite{usadeleq}. Thus, our model reproduces, in the diffusive
limit, the results of Refs.\cite{BVE_inverse,BVE_screening} for
the induced magnetization.

An intermediate situation between the purely ballistic and the
diffusive limits can be analyzed by introducing a parameter
$\alpha$ controlling the proportion of the hopping term which is
random, i.e. $t_B = \alpha t_{SF}$ and $\delta t = (1 - \alpha)
t_{SF}$ in such a way that $\alpha = 1$ ($\alpha=0$) corresponds
to the ballistic (diffusive) case. In a general intermediate case
the self-energy can be written as a sum of a ballistic
$\Sch_{S(F)}^b$ and a diffusive $\Sch_{S(F)}^d$ contribution,
calculated according to Eqs. (\ref{Sigma_b}) and (\ref{Sigma_d}).



\section{Thin S/F bilayer}

Let us consider two thin S/F layers, such that the unperturbed
Green functions are given by Eq. (\ref{gunp}). Our aim is to
calculate the magnetic moment $M$ induced in the superconductor,
which is given by
\[
M=\mu_BT\sum_{\omega} {\rm
Tr}\left[\langle\hat{G}_{S+}(k,\omega)-\hat{G}_{S-}(k,\omega)\rangle\right]\;.
\]
As it was mentioned above we assume a linear dispersion relation
(flat bands). This means that
\[
\langle G\rangle=\nu\int d\epsilon G\; ,
\]
where $\nu$ is the normal density of states at the Fermi level.
Notice that in the wide-band approximation that we are using here
the magnetization arises  from changes in the density of states
close to the Fermi level.

The exact normal Green functions $\hat{G}_{S,F\pm}$ for spin up
and spin down can be written in the more general case as (see Eq.(\ref{dyson}))
\begin{eqnarray}
  \Ght_{S\pm} &=& \left[\ghts^{-1}-\Shts^{b}-\Shts^{d}\right]^{-1} \\
  \Ght_{F\pm} &=&
  \left[\ghtf^{-1}-\Shtf^{b}-\Shtf^{d}\right]^{-1}.
\end{eqnarray}
The self-energies $\hat{\Sigma}^{b,d}$ correspond to the terms
defined in Eqs.(\ref{Sigma_b}-\ref{Sigma_d}). We thus obtain the
following set of coupled equations
\begin{eqnarray}
  <G_{S\pm}> &=& -i\langle \frac{D_\pm\Omega_\pm-it_B^2\omega_\pm}{t_B^4+2t_B^2(\omega_\pm\Omega_\pm-\epsilon_k^2)+D_\pm(\Omega_\pm^2+\Delta_\pm^2+\epsilon_k^2)}\rangle \label{g1}\\
  <F_{S\pm}> &=& \langle \frac{D_\pm\Delta_\pm}{t_B^4+2t_B^2(\omega_\pm\Omega_\pm-\epsilon_k^2)+D_\pm(\Omega_\pm^2+\Delta_\pm^2+\epsilon_k^2)}\rangle\\
   <G_{F\pm}>&=&-i\langle\frac{D_s\Lambda_\pm+\omega t_B^2}{t_B^4+2t_B^2(\omega\Lambda_\pm-\epsilon_k^2\mp \Delta\delta t^2<F_{S\pm}>)+D_s(\Lambda_\pm^2+\epsilon_k^2+\delta t^4<F_{S\pm}>)}\rangle  \\
  <F_{F\pm}> &=&\langle \frac{\mp t_B^2\Delta+\delta t^2D_S<F_{S\pm}>}{t_B^4+2t_B^2(\omega\Lambda_\pm-\epsilon_k^2\mp \Delta\delta t^2<F_{S\pm}>)+D_s(\Lambda_\pm^2+\epsilon_k^2+\delta
  t^4<F_{S\pm}>)}\rangle\label{g2}
\end{eqnarray}
where we have defined the following quantities:
$D_\pm=\omega_\pm^2+\epsilon_k^2$;
$D_S=\omega^2+\epsilon_k^2+\Delta^2$; $\Omega_\pm=\omega+i\delta
t^2<G_{F\pm}>$; $\Delta_\pm=\pm \Delta+\delta t^2<F_{F\pm}>$ and
$\Lambda_\pm=\omega_\pm+i\delta t^2<F_{F\pm}>$. This set of
equations must be solved in a self-consistent way. However for the
pure ballistic case, when $\delta t=0$, the equations in both
layers are decoupled. For example it is easy to show that
\begin{equation}\label{Gupsol}
G_{S+}=-\frac{i\omega(\omega_+^2+\epsilon_k^2)+it^2\omega_+}{(\epsilon_k^2-\epsilon_1^2)(\epsilon_k^2-\epsilon_2^2)}\;,
\end{equation}
where
\begin{equation}\label{xis}
\epsilon_{1,2}=\sqrt{\frac{-\omega^2-\omega_+^2-\Delta^2+2t^2\pm
\sqrt{(\omega^2-\omega_+^2+\Delta^2)^2-4t^2\left[(\omega+\omega_+)^2+\Delta^2\right]}}{2}}
\end{equation}
For $G_{S-}$ we obtain a similar expression substituting
$\omega_+$ by $\omega_-$. Now the  integration over $\epsilon$ can be
performed by closing the integration path in the upper half-plane
and using the residue theorem.
\begin{figure}[b]
 \includegraphics{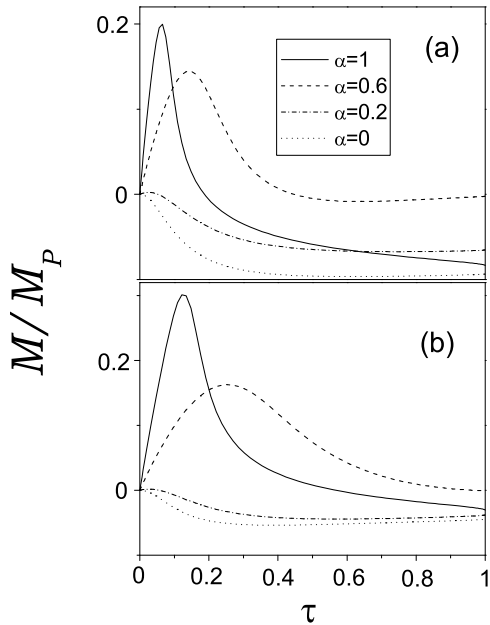}
\caption{ Induced magnetization $M/M_P$ as a function of the
transmission $\tau$ for (a) $h=0.2$ and (b) $h=0.4$ and different
values of $\alpha$. We have chosen $\Delta=0.1$\label{thinsf} and
$T=10^{-4}$. All energies are given in units of $t_S=t_F$.}
\end{figure}

In Fig. \ref{thinsf} we show the dependence of the  induced
magnetic moment on the interface transparency, for very low
temperature ($T\ll\Delta$) and different values of the parameter
$\alpha$ and of the exchange field $h$. $M$ is given in units of
the Pauli paramagnetic moment $M_P$ defined as
\[
M_P=2\mu_B\nu h\; .
\]
The interface transparency is characterized by the transmission
coefficient $\tau$ , which within this model is given by $\tau =
4(t_{SF})^2/(1 + (t_{SF})^2)^2$. Throughout this paper all
energies are given in units of $t_S=t_F$.

According to Fig. \ref{thinsf} in the pure ballistic case
($\alpha=1$) the dependence $M(\tau)$ has a nonmonotonic behavior:
in accordance with the results obtained in Ref. \cite{buzdin05}
for low transparency, the induced magnetic moment is positive.
However, at some $\tau$ (when $t_{SF}$ is of the order of $h$) $M$
changes to negative values. If the exchange field is large enough
this change of sign would never take place and the magnetization
remains positive for all values of $\tau$.

Figure \ref{thinsf} also shows that by decreasing $\alpha$, i.e.
by increasing the disorder at the interface, the curves become
flatter.  In the limiting case of a diffusive interface, i.e. when
$\alpha=0$, our model leads to the same result as the one obtained
using the quasiclassical approach in Ref.\cite{BVE_inverse}. This
is not surprising since, as it is shown in the appendix, our model
in the diffusive limit ($\alpha=0$) leads to the Usadel equation
and therefore all conclusions of Ref.\cite{BVE_inverse} also hold
here. In particular the physical picture given in Ref.
\cite{BVE_inverse} of Cooper pairs sharing their electrons between
both the S and the F layer, may explain the negative induced
magnetic moment. From Eqs.(\ref{g1}-\ref{g2}) one can easily check
that in the Usadel limit ($\alpha=0$) and for low transparencies
the first correction to the density of states, in both layers, is
proportional to $\delta t^4$, i.e. to the square of the interface
transparency. This indicates that in this limit the inverse
proximity effect arises from higher order processes requiring the
induction of a non-vanishing paring amplitude in the ferromagnet.
In fact, the Green functions in the F region have a BCS-like form
with an induced minigap in the DoS { \cite{zareyan,BVE_inverse}}.
We are thus dealing with two coupled condensates with Cooper pairs
sharing the electrons between the S and the F layer. This coupling
tends to align the spins of both regions in opposite direction.

\begin{figure}[h]
\includegraphics[angle=-90,scale=0.5]{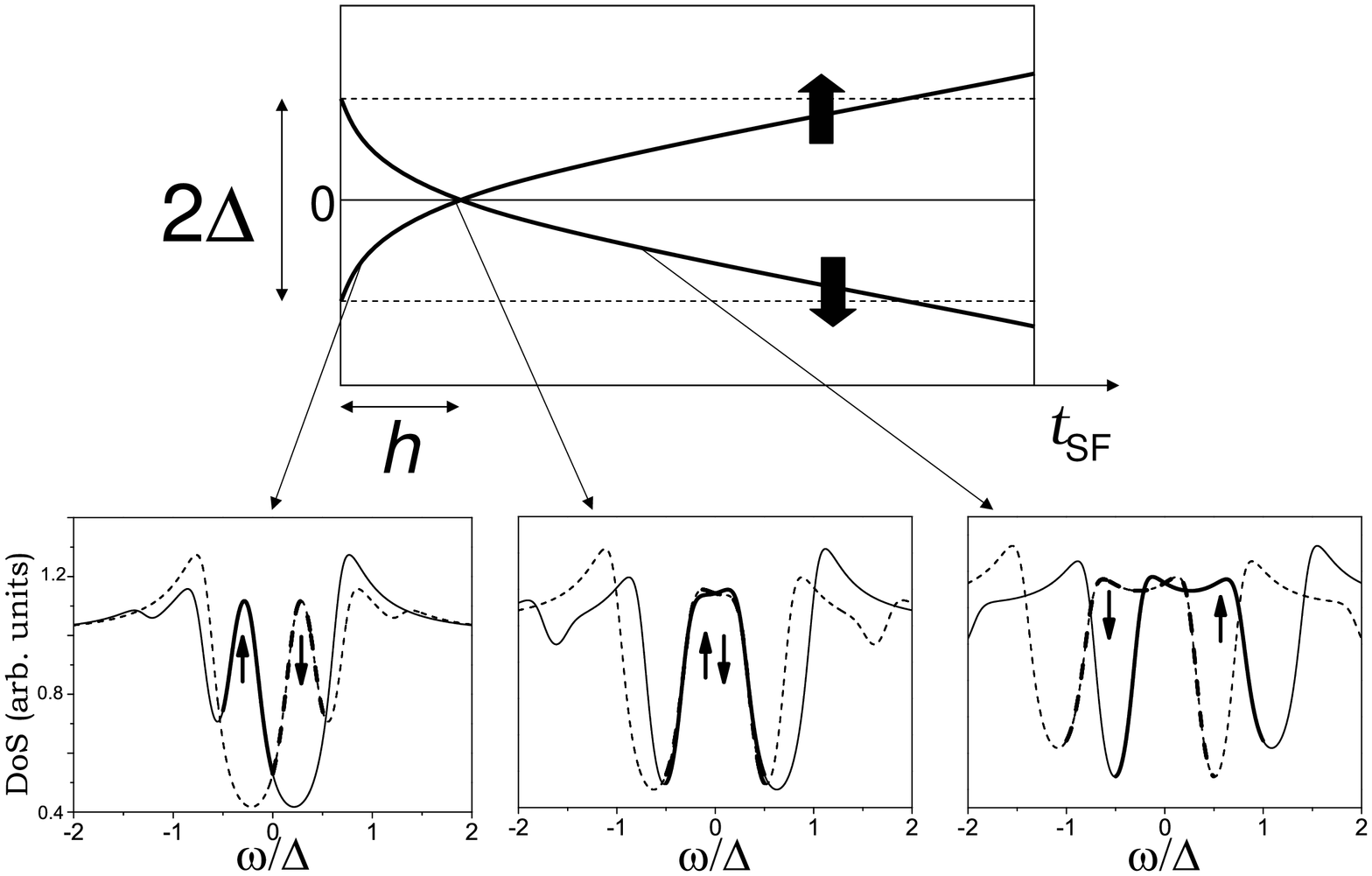}
\caption{Sketch of the position of the Andreev states in the
spectral density for the majority ($\uparrow$) and minority
($\downarrow$) spins as a function of the hopping parameter
$t_{SF}$. The three panels at the bottom  show the DoS of the
spin-up (solid line) and spin-down (dashed line) electrons for
three different values of $t_{SF}$. At low transparencies and
below the Fermi level there is a peak in the spin-up channel (left
panel). Increasing the value of $t_{SF}$ this peak moves to the
right while the peak in the spin-down channel moves to the left.
At some $t_{SF}$ of the order of $h$ both peaks coincide (middle
 panel). Further increase of $t_{SF}$ leads to a spin-down peak below the Fermi level (right panel).
 \label{andreev}}
\end{figure}

It would be interesting to find a physical argument for explaining
the observation of a positive magnetization in the ballistic
regime. In this case one can check from Eqs. (\ref{g1}-\ref{g2})
that if $\alpha=1$  and  for low transparency, the first
correction to the induced magnetic moment is of  order $t_{SF}^2$
and is positive ({\it cf.} Ref. \cite{buzdin05}). However, when
$\tau$ becomes larger the induced magnetization reaches a maximum
and then decreases down to negative values. One can understand
this behavior in terms of Andreev states induced in the DoS. In
Fig.\ref{andreev} we sketch the  position of these states in the
spectral density. For values of $t_{SF}$ smaller than $h$ the
state corresponding to the majority spin is located below the
Fermi energy and thus, the induced magnetization has a positive
sign. The corresponding density of states is shown in the inset of
Fig. \ref{andreev}, where a peak in the spin-up channel (solid
line) can be identified. By increasing the interface transparency
and when $t_{SF}\sim h$  the Andreev states for both spin
orientations eventually cross. A further increase of $\tau$ leads
to a negative magnetization, since now the Andreev state below the
Fermi level corresponds to the spin-down orientation (the dashed
curve in the right inset of Fig. \ref{andreev} shows the DoS for
the spin-down electrons).

The S/F bilayer  of atomic thickness analyzed so far should not be
considered as a mere idealized example but tt could also describe
the situation in real hight $T_C$ materials consisting of
alternating magnetic and superconducting layers.

\section{S/F layers of arbitrary thickness}

Let us consider a superconductor of thickness $d_S\gg \xi_S$
(where $\xi_S=2at_S/\pi\Delta$ is the coherence length) in contact
with a ferromagnet of arbitrary thickness which is the situation
in most experiments \cite{garifullin,stahn}. This case can be
analyzed theoretically considering the superconductor as a
semi-infinite system. We shall first consider the ballistic regime
in which electrons are specularly reflected ($k$ is conserved) at
the outer surface of the ferromagnet.
\begin{figure}[h]
\includegraphics[scale=1]{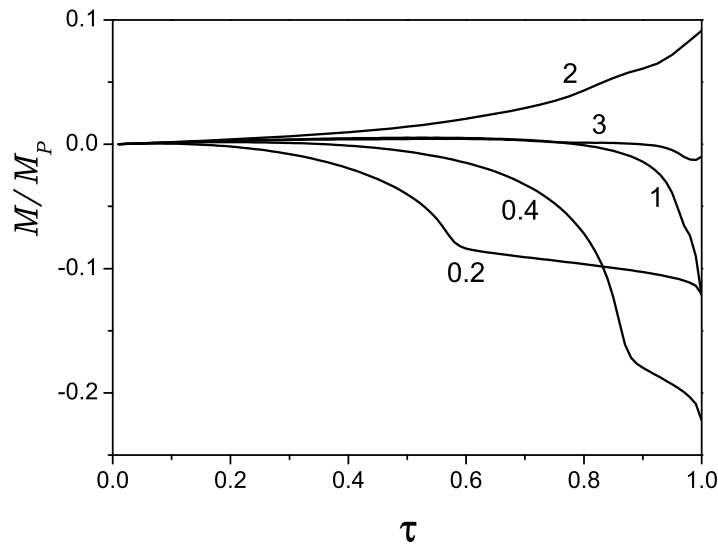}
 \caption{The induced magnetization $M/M_P$ at the S side of the interface as a
function of the transmission coefficient $\tau$ for different
values of the F-layer thickness $d_F$.  The latter is given in
units of $\pi\xi_F$. The S layer is assumed to be  semi-infinite.
We have chosen $h=0.2$, $\Delta=0.1$ and T=0. \label{difnf}.}
\end{figure}
In Fig. \ref{difnf} we show the magnetic moment induced in the
superconductor at the S/F interface as a function of the
transmission coefficient  $\tau$ and for different thickness $d_F$
of the F layer.  In the tunnelling limit the effect is very weak
as the induced magnetization is proportional to $t_{SF}^4$.
However, in the region of high transparency an appreciable
magnetic moment can be induced in the superconductor. {As it can
be observed in Fig. \ref{difnf}, its sign depends on the ratio
between the thickness of the F layer and the ferromagnetic length
defined as $\xi_F=at_F/\pi h$: while for thin layers ($d_F \le
\pi\xi_F$) is negative, by increasing the F layer thickness it
becomes positive reaching a maximum for $d_F \sim 2\pi\xi_F$. For
$d_F \gg \xi_F$ the induced magnetization tends to zero. Since the
structure is purely ballistic we expect that interference effects
take place leading to a nonmonotonic behavior. The change of sign
can be understood within the description of Andreev bound states
similar to the one depicted in the previous section. According to
the calculations of Ref. \cite{vecino}  if $\tau$ is of the order
of unity, the crossing of the Andreev peaks associated to the
majority and minority spins, and thus  the change of sign of the
induced magnetization, occurs at $d_{F0}\sim4\xi_F$}. This is in
agreement with the present results.

\begin{figure}[h]
 \includegraphics[scale=1]{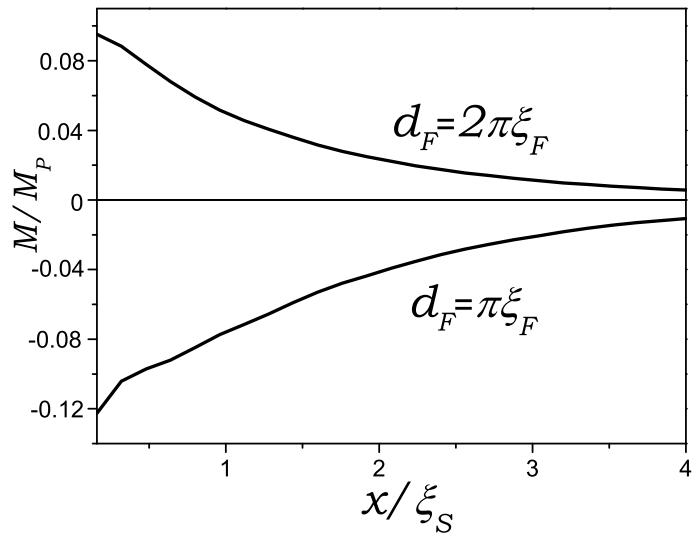}
 \caption{The spatial dependence of the magnetization induced in a semi-infinite S layer in contact
 with a F layer of thickness $\pi\xi_F$ and $2\pi\xi_F$.
 We have chosen $h=0.2$ and $\Delta=0.1$ \label{spatialdp}.}
\end{figure}

\begin{figure}[h]
\includegraphics[scale=1]{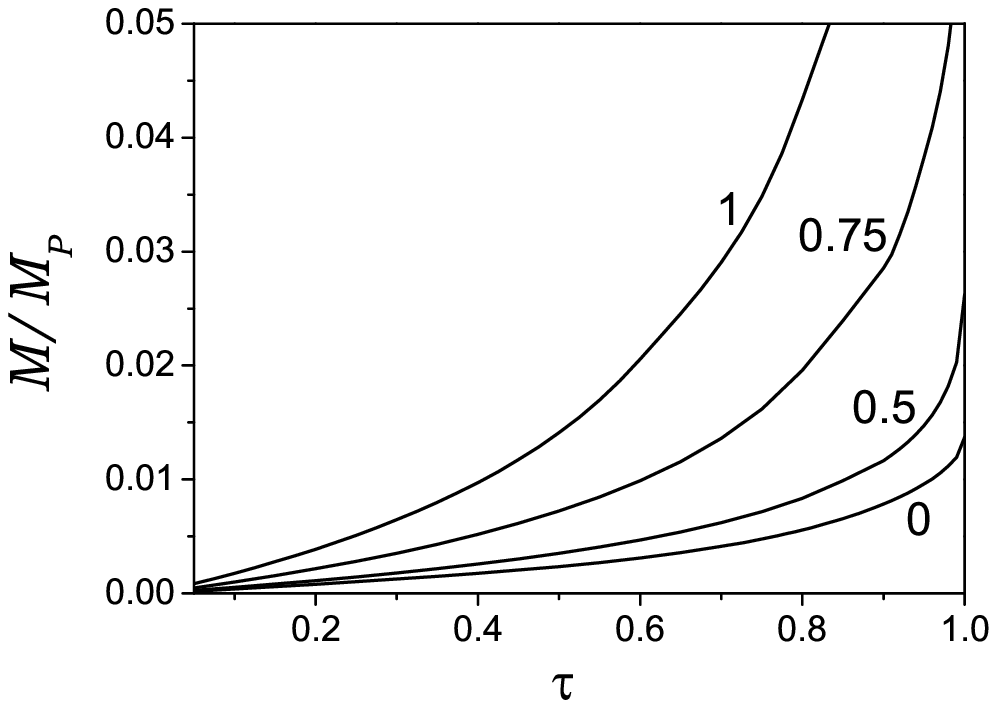}
\caption{Magnetization in the superconductor as a function of
$\tau$ for different type of interfaces: $\alpha=1$ corresponds to
the ballistic while $\alpha=0$ to the diffusive one. The other
values chosen for $\alpha$ are 0.5 and 0.75. Here $d_F=2\pi\xi_F$,
$h=0.2$, $\Delta=0.1$ and $T=0$. \label{mvsalpha}}
\end{figure}
{A natural question that arises at this point concerns the
penetration length of the induced magnetization}. In Fig.
\ref{spatialdp} we show the spatial dependence of $M$ in the
superconductor for two different thicknesses of the F layer and
the same values of $h$ and $\Delta$ as in  the previous figure. We
have chosen $d_F=\pi\xi_F$ and $2\pi\xi_F$ in order to show that
whatever the sign of the induced magnetization is, the penetration
length is of the order $\xi_S$. This demonstrates that the effect
described here is due to the superconducting proximity effect and
it is not just a magnetization leakage over atomic distances as in
Ref. \cite{halterman_04}.

Finally we analyze the case in which the two layers are connected
by a disordered interface. In Fig. \ref{mvsalpha} we show the
dependence of $M/M_P$ on the transmission for different values of
the parameter $\alpha$ and $d_F=2\pi\xi_F$. {We observe that by
increasing $\alpha$, the induced magnetization is reduced.
However, even in the case of a fully disordered interface, a
finite magnetization of the  order of $10^{-2}M_P$ for large
transparency is obtained, demonstrating that the effect is robust
against disorder.}

\section{Conclusions}
{We have studied the inverse proximity effect in S/F structures
with arbitrary thickness, type of interfaces and degree of
disorder. For this purpose we have presented a microscopic model
which combines a model Hamiltonian with elements of the
quasiclassical theory. We have shown that the sign and magnitude
of the magnetic moment induced in the superconductor is highly
dependent on the system parameters. In the diffusive limit, where
our model leads to a discretized version of the well known Usadel
equation, the induced magnetization in S and the one in F have
opposite directions, in agreement with Ref. \cite{BVE_inverse}.
However, we have shown that the induced magnetization can exhibit
either sign when the system deviates from the diffusive regime.}

{We have first considered thin layers and shown that in the purely
ballistic case and for low transparency of the interface the
induced magnetization is positive.  By increasing the transmission
the magnetization reaches a maximum and then decreases to negative
values. The change of sign takes place when the coupling parameter
between the layers is of the order of the exchange field. This
behavior can be understood in terms of localized Andreev states
and their evolution when varying the coupling strength. The
crossover between the ballistic and diffusive limits was also
analyzed.}

{Finally we have studied a bulk superconductor attached to a
ferromagnetic layer of arbitrary thickness $d_F$.  We have shown
that for high interface transparency  and small $d_F$ the induced
magnetization $M$ is negative. For increasing thickness and when
$d_F$ is of the order of $\xi_F$ a change of sign in the induced
magnetization takes place. In both cases (positive or negative
sign of the magnetization) $M$ penetrates into the superconductor
over a distance comparable to the superconducting coherence length
demonstrating that  the effect is long-ranged. }

{It is worth mentioning that band structure effects as well as a
$d$-wave symmetry for the superconducting order parameter can be
straightforwardly included in our Hamiltonian approach. This will
be the object of future works.}

{We expect that experimental techniques such as neutron
reflectometry or muon spin rotation used recently in Refs.
\cite{stahn,muon} may help to confirm our predictions. }

\acknowledgements
 {This work was supported by the European Union
through the DIENOW Research Training Network}

\appendix

\section{The self-energy term for different types of interfaces}\label{appA}
Let us consider the purely ballistic case which occurs when the
mean free path of electrons is much larger than the characteristic
dimension of the system.
 As we are considering layered structures, with translational invariance parallel to the interface
($y-z$ plane in Fig. \ref{geometry}), it is convenient to perform
a change of basis from the local one used in writing  the model
Hamiltonian (\ref{hs}-\ref{hsf}) into a mixed basis of Bloch waves
in the direction parallel to the interface and localized states in
the perpendicular direction. Thus, we subdivide the layers into
planes parallel to the interface, separated by a lattice spacing
$a$ \footnote{In principle the thickness $a$ is of the order of
the Fermi wave length $\lambda_F$.}. We  label these planes by
integer numbers $n$.  The new basis is given by the state vectors
$|\vec{k},n>= \sum_{j\in n} e^{i\vec{k} \vec{R}_j} |j>$,
 where $\vec{k}$ is the momentum parallel to the plane of the junction, while
$\vec{R}_j$ is the position of the site $j$ within the plane. A
similar model was considered in Ref. \cite{melin04}.

After changing the basis, the field operators in
Eqs.(\ref{hs}-\ref{hsf}) are now labelled by $\vec{k}$ and $n$. We
assume that the superconducting order parameter is homogeneous in
the superconductor and therefore $\Delta_k=\Delta$. In principle
the dispersion relation $\epsilon_k$ may be non-linear. However in
the present work we assume that all energies involved in the
problem ($\Delta$, $h$, etc.) are much smaller than the Fermi
energy and thus we can assume a linear and isotropic {dispersion
relation, i.e. $\epsilon_k \sim v_F(k-k_F)$}, where $v_F$ is the
Fermi velocity.  A generalization of this model beyond this
quasiclassical approximation and including  a more realistic band
structure could be implemented straightforwardly.

Since the system is purely ballistic, the parallel momentum
$\vec{k}$ is conserved across the interface. This means that the
hopping matrices ($\check{T}_S$, $\check{T}_F$, $\check{T}_{SF}$)
in the new basis are diagonal, i.e.
\[< \vec{k}^{\prime}n|\check{T}|\vec{k}n+1>=\delta_{kk^{\prime}}t_k.\]
For simplicity we shall assume that the lattice model is such that
$t_k = \sum_{<i,j>} e^{i\vec{k}\vec{R}_{ij}} t_{ij}$ do not depend
on $\vec{k}$. {This is consistent with the approximation of a
linear dispersion relation introduced above.} Thus, the
self-energy term in the Dyson equation Eq.(\ref{dyson}) is given
by
\[
    \Sch_{S(F)}^b(k)=t_{SF}^2\tau_3\check{g}_{F(S)}(k)\tau_3\;,
\]

In order to calculate the interface Green functions $\gch_{S(F)}$
of the uncoupled regions one needs to know the functions
$\gch_{S(F)}^{(0)}$ of the uncoupled planes. These are given by
\cite{BVE_josephson}
\begin{equation}\label{gunp}
\gch_{S(F)}^{(0)}=\left(\begin{array}{cc} \hat{g}_{S(F)+}^{(0)}+\hat{f}_{S(F)+}^{(0)} & 0\\
0 & \hat{g}_{S(F)-}^{(0)}+\hat{f}_{S(F)-}^{(0)}
\end{array}\right)\; ,
\end{equation}
where the normal and anomalous components,
$\hat{g}_{S(F)\pm}^{(0)}$ and $\hat{f}_{S(F)\pm}^{(0)}$,  for spin
up ("+") and down("-") are known and given by
\begin{eqnarray}
\ghts^{(0)} &=&\frac{-i\omega-\epsilon_k\tau_3}{\omega^2+\Delta^2+\epsilon_k^2}\label{gs}\\
\fhts^{(0)}
&=&\frac{\pm\Delta\tau_1}{\omega^2+\Delta^2+\epsilon_k^2}\label{fs}\\
\ghtf^{(0)}&=&-\frac{i\omega_{\pm}-\epsilon_k\tau_3}{\omega_{\pm}^2+\epsilon_k^2}\label{gf}\;
,
\end{eqnarray}
where $\tau_i$ are the Pauli matrices in particle-hole space,
$\omega_{\pm}=\omega\mp ih$, and $\omega=\pi T(2n+1)$ are the
Matsubara frequencies. It is clear that the anomalous component
$\fhtf^{(0)}$ of the unperturbed Green function equals zero in the
F layer.

The purely ballistic case is unrealistic for describing many
experimental situations. Even when the uncoupled S and F layers
are clean enough, some degree of disorder at the interface is in
general unavoidable. We will show that  the inverse proximity
effect  strongly depends on the type of interface and the degree
of disorder.

\begin{figure}[h]
\includegraphics[angle=-90, scale=0.5]{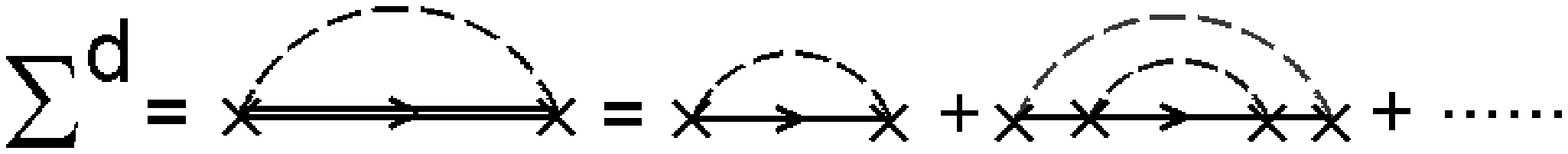}
 \caption{The self-energy  within the self-consistent Born approximation.The thick lines in
this figure correspond to the fully dressed Green function, the
crosses represent the hopping processes and the dashed semicircle
indicates averaging over the parallel momentum $k$ \label{diag}.}
\end{figure}

We start by analyzing the case of a disordered interface between
the S and the F layer. In this case the layers are connected by
random hopping terms $\delta t_{ij}$. These random terms are
assumed to follow a Gaussian locally correlated distribution, i.e.
\begin{equation}
\overline{\delta t_{ij} \delta t_{lm}}  = \delta t^2 \delta_{im}
\delta_{jl}. \label{dist}
\end{equation}
The observable quantities are calculated by averaging over
disorder configurations. This averaging procedure restores the
translational invariance in such a way that the averaged Green
functions satisfy a Dyson equation like in the ballistic case (Eq.
(\ref{dyson})) but with a different self-energy. In order to
obtain an approximate self-energy for the disordered case we use
the self-consistent Born approximation, whose diagrammatic
representation is sketched in Fig. \ref{diag}
\cite{bauer_schep,brataas_bauer}. The thick lines in this figure
correspond to the fully dressed Green function, the crosses
represent the hopping processes and the dashed semicircle
indicates averaging over the parallel momentum $k$. The expression
for the self-energy is then given by
\[
\Sch_{S(F)}^d=\delta
 t^2\tau_3\langle \hat{G}_{F(S)\pm}\rangle\tau_3\; ,
\]
where $<...>$ denotes integration over parallel momentum $k$.

\section{Comparison with quasiclassics}\label{appB}

 According to our model  a diffusive
metal can be also described by a random coupling $\delta t_{ij}$
between the layers with a distribution given by Eq. (\ref{dist}).
The Green function at the plane $n$ is determined by the Dyson
equation
\begin{equation}\label{n}
    G_n=g_n+g_n\delta t^2<\tau_3G_{n+1}\tau_3>G_{n}+g_n\delta t^2<\tau_3G_{n-1}\tau_3>G_{n}\; ,
\end{equation}
where $g_n$ denotes the Green function for the uncoupled plane
(cf. Eqs. (\ref{gs}-\ref{gf})) and $<...>$ the average over
momentum. Multiplying this equation by $g_n^{-1}$ from the left
and by $\tau_3$ from the right and subtracting the conjugated
equation multiplied by $g_n^{-1}$ from the right and by $\tau_3$
from the left one obtains
\begin{equation}
\left[\tau_3g_n^{-1},\tilde{G}_n\right]=\left[\Sigma_{n+1},\tilde{G}_n\right]+\left[\Sigma_{n-1},\tilde{G}_n\right]\label{a1}
\end{equation}
where $\tilde{G}_n=G_n\tau_3$, $\Sigma_n=t^2<\tilde{G}_n>$ and
$[,]$ denotes a commutator.

As the dependence of $g_n^{-1}$ on the momentum only appears in
the term proportional to $\tau_3$ which vanishes when subtracting
the two Dyson equations, the integration over momentum can be done
easily and is equivalent to substitute $\tilde{G}$ by
$<\tilde{G}>$. One obtains
\begin{equation}
\left[\tau_3g_n^{-1},\tilde{G}_n\right]=
t^2\left(<\tilde{G}_{n+1}>+<\tilde{G}_{n-1}>\right)<\tilde{G}_{n}>-
t^2<\tilde{G}_{n}>\left(<\tilde{G}_{n+1}>+<\tilde{G}_{n-1}>\right)
\label{a11}
\end{equation}

Note that by subtracting Eq. (\ref{n}) from its conjugate the
inhomogeneous term also cancels out. Thus, we need a second
equation in order to determine the Green functions. As in the
derivation of the quasiclassical equations, in the wide-band
approximation considered in this work, the Green functions
satisfies the normalization condition $<\tilde{G}>^2=c$, where $c$
is a constant \cite{eilenberger}.

One can associate the terms in  the r.h.s of Eq. (\ref{a11}) with
the second spatial derivative of a continuous model by means of
the identification
\begin{equation}
\frac{<\tilde{G}_{n+1}>+<\tilde{G}_{n-1}>-2<\tilde{G}_n>}{a^2}\equiv\partial_{xx}^2{\cal
G}\label{a2}
\end{equation}
From the normalization condition follows that the second
derivative of ${\cal G}^2$ vanishes and therefore
\begin{equation}
2\partial_x({\cal G}\partial_x{\cal G})={\cal G}
.\partial_{xx}^2{\cal G}-\partial_{xx}^2{\cal G} .{\cal
G}\label{a3}
\end{equation}
Thus, using Eqs. (\ref{a11}-\ref{a3}) we obtain following equation
for ${\cal G}$
\[ \left[\tau_3g^{-1},{\cal G}
\right]=-2t^2a^2\nu\partial_x({\cal G}
\partial_x {\cal G} )
\] Identifying the factor $t^2a^2\nu$ with the diffusion
coefficient one sees that  this expression is equivalent to the
well known Usadel equation  \cite{usadeleq}.


\end{document}